\documentstyle[prl,aps,psfig,twocolumn]{revtex} % COMMENT OUT IF

\begin{document}
\draft
%                            % COMMENT OUT NEXT LINE IF PREPRINT
\twocolumn[\hsize\textwidth\columnwidth\hsize\csname
@twocolumnfalse\endcsname
\title{Universality and Crossover of Directed Polymers
and Growing Surfaces}
\author{Paolo De Los Rios}
\address{Institut de Physique Th\'eorique, 
Universit\'e de Fribourg, CH-1700, Fribourg, Switzerland.}
\date{\today}
\maketitle

\begin{abstract}
We study KPZ surfaces on Euclidean lattices
and directed polymers on hierarchical lattices 
subject to different distributions
of disorder, showing that universality holds, at odds with
recent results on Euclidean lattices. 
Moreover, we find the presence of a slow (power-law) crossover toward the 
universal values of the exponents and verify that the exponent governing such
crossover is universal too.
In the limit of a $1+\epsilon$ dimensional system we
obtain both
numerically
and analytically that the crossover exponent is $1/2$.
\end{abstract}
]
\narrowtext
 
The problem of Directed Polymers in Random Media (DPRM) \cite{HH85+} 
has attracted
much attention in the last ten years, both as a paradigm in the area of
disordered systems and for the richness of its connections with other
systems, in particular noisy surface growth governed 
by the Kardar-Parisi-Zhang (KPZ) equation\cite{KPZ86+,HHZ95}.
Both problems show space and time scaling behavior, 
and the connection between the two manifests through
a correspondence between their exponents.
Within the KPZ context, it has been recently proposed that 
the model could be non-universal\cite{NS97}: The exponents characterizing
the surface growth depend on the details of the driving noise,
at least for a substrate dimension $d \ge 2$.
These results have been obtained {\it via} a lattice formulation
of the KPZ equation where the strong coupling limit (which is the
non trivial regime of surface growth) is shown to be completely 
equivalent to the ground state problem of DPRM's. 
Thus, we can expect non-universality in the DPRM context as well.
It is therefore interesting
to re-examine this issue both within the KPZ context and for DPRM's.

Before proceeding further, it is useful to recall the definition of the
exponents within the two contexts, and their relations.
Starting from a flat substrate of characteristic linear size $L$, the roughness 
of a KPZ surface grows initially as $W(t,L) \sim t^\beta$; at longer times
it saturates, and it scales with $L$ as $W(t,L) \sim L^\chi$.
The characteristic time $\tau$ between the two regimes scales
with the size of the system as $\tau \sim L^z$. 
These exponents are not independent: the relation (rooted
in the Galilean invariance of the KPZ equation)
$\chi + z = 2$ holds in every dimension\cite{KPZ86+,HHZ95}.
Moreover, consistency imposes $\chi = z \beta$.
Therefore there is just a single independent exponent. 
In the language of DPRM's, the exponent $\beta$ governs the
fluctuations of the ground state energy,
$\Delta E = \sqrt{<(E_{GS} - <E_{GS}>)^2>} \sim t^\omega$ with
$\omega = \beta$ (here we use the greek alphabet letters commonly used in the
literature); the transverse wandering fluctuations of the ground
state polymer are governed by the exponent $\zeta = 1/z$,
$\Delta l \sim t^\zeta$. The relation $\omega = 2 \zeta -1$ holds
in every dimension (it is related to $\chi + z = 2$).
%In what follows we are interested in the
%ground-state properties of DPRM's. In particular their
%behavior can be characterized by $2$ exponents, $\omega$ and 
%$\zeta$. The former rules the scaling behavior of the ground-state energy
%fluctuations, $\langle \Delta x \rangle 
%= \sqrt{\langle x^2 \rangle - \langle x \rangle^2} \sim
%t^\omega$, where the angular brackets indicate ensemble averages, and $t$
%is the length of the lattice; the exponent $\zeta$ tells how
%the transverse wandering fluctuations $\Delta l$ 
%of the ground state configuration scale
%with the length of the lattice, $\langle \Delta l \rangle \sim t^\zeta$.
%These exponents are not independent: the relation (rooted
%in the Galilean invariance of the equivalent KPZ equation)
%$\omega = 2 \zeta -1$ holds in every dimension\cite{KPZ86+,HHZ95}. 

The exponents are known exactly only
for $d=1$: $\beta=1/3$ and $z=3/2$ ($\omega = 1/3$, $\zeta = 2/3$). 
At present there are no exact solutions for $d \ge 2$, and
our knowledge of the exponents relies on numerical simulations.
Yet, the situation is not completely clear even numerically:
As it has been pointed out recently~\cite{NS97}, in the context
of surface growth (KPZ equation) different
distributions of disorder seem to give different values of the exponents.
This result has been interpreted as a case of non-universality.
Indeed, using a distribution of the energies as
\begin{equation}
p(x) = \frac{(1-\alpha)}{2} (1-|x|)^{-\alpha},\;\; x \in (-1,1),\;\;
\alpha > -1\;\;,
\label{swift distribution}
\end{equation}
in \cite{NS97} it was found that the values of the exponents depended
on the value of $\alpha$ in (\ref{swift distribution}).
In \cite{NS97} the exponents were calculated only for $d \ge 2$, assuming
that universality holds in $d=1$, where the exponents are exactly known
(it is worth mentioning that the exact knowledge of the exponents
is based on taking a Gaussian distribution of disorder).

\begin{figure}
\centerline{\psfig{file=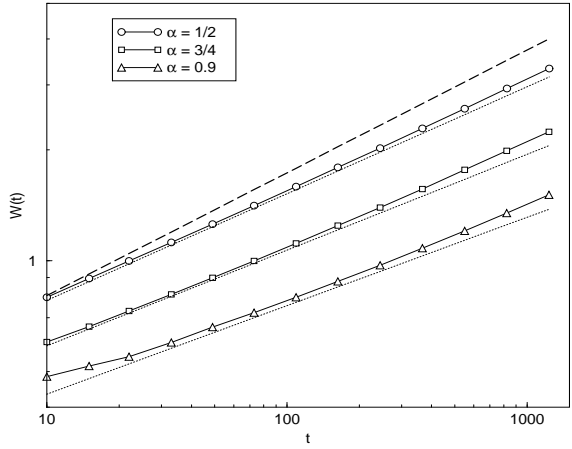,width=7.5cm,height=6cm,angle=270}} 
\caption{Roughness of KPZ surfaces for exponents $\alpha =0.5, 0.75, 0.9$
in Eq.(1).  The dashed line represents a power-law with exponent $\beta = 1/3$,
the dotted lines represent temptative fits. Each numerical point
is the average from $10000$ independent disorder realizations.} 
\label{Fig: fig1}
\end{figure}

\begin{figure}
\centerline{\psfig{file=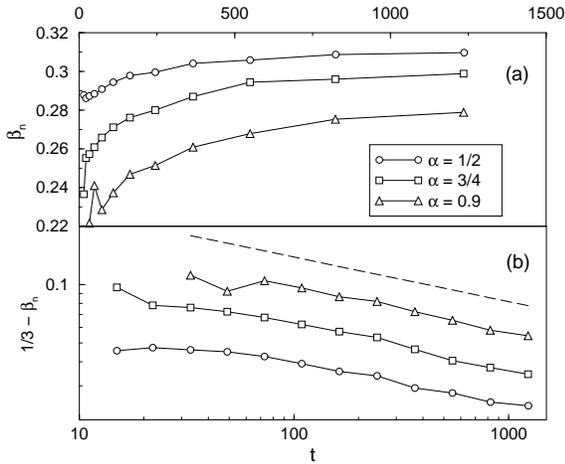,width=7.5cm,angle=0}} 
\caption{(a) Running exponents for the roughness as in Fig.1. (b) Log-log
plot of the difference of the running exponents from the theoretical value
$1/3$. The dashed line represents a power-law with exponent $0.23$.} 
\label{Fig: fig2}
\end{figure}

To settle this problem, we have performed simulations of surface growth
on $d=1$ Euclidean lattices. We measure the $\beta$ exponents
ruling the growth in time of the roughness $W(t)$ starting from
a flat substrate (the other exponents can be obtained from the
above mentioned relations). We used three different
values of $\alpha$, namely $\alpha=0.5\;,0.75\;, 0.9$ (in
each case we also verified that we are still far from 
saturation).
As it can be seen from Fig.\ref{Fig: fig1},
a power law with exponent $1/3$ (the theoretical one)
is not suited to fit the numerical power laws obtained.
A na\"if fit would give instead $\beta=0.28\;,0.26\;,0.24$ respectively.
Even fitting an exponent on the last four points with $\alpha=0.5$
would give $\beta=0.30$ (results for the most commonly used case, 
the uniform distribution with $\alpha=0$, 
are not given since the fitted exponent is $\beta=0.31$
and the crossover is less evident). To understand whether this 
is a case of non-universality or of crossover,
we also analyze the running exponents (see
Fig.\ref{Fig: fig2}a). If the numerical points shown in Fig.\ref{Fig: fig1}
are taken at fixed time ratios $t_{n+1}/t_n = r$,
then the running exponents are defined as
\begin{equation}
\beta_n = \log_r \frac{W(t_{n+1})}{W(t_n)}
\label{running in time}
\end{equation}
where, in the present case, $r=3/2$. We find that the running exponents 
approach their universal value $\beta=1/3$. In a log-log plot 
(Fig.\ref{Fig: fig2}b) it is
then easy to see that such an approach is ruled by a power-law with
a universal (independent of $\alpha$) exponent $\gamma=0.23 \pm 0.02$.

The results obtained are amenable of different interpretations: either
the $d=1$ case is non universal too, against naive expectations,
or it is indeed affected by extremely slow crossovers. 
The analysis of the running exponents seems to sustain the latter
hypothesis. The behavior in higher dimensions 
is unfortunately much more difficult to extract. Indeed,
we have also performed simulations in $d=2$, but the results,
although similar, are
somehow less conclusive due to numerical limitations: in order
to get reliable statistics in $d=1$, each numerical point is averaged
over $10000$ independent realizations. Moreover, universality in the crossover
is evident only for times $\ge 100$. Matching both conditions
for  $d \ge 2$ goes beyond the capabilities of our present computing
facilities.  

To investigate the 
high dimensional behavior of this crossover we resort then
to analytical and numerical calculations on hierarchical lattices,
where the same behavior emerges. 
%
%Hierarchical lattices have been widely used in recent years
%in many contexts. In particular they have proven to be useful
%in many polymer-related problems, since many analytical or semi-analytical 
%calculations can be performed, and exponents can be extracted with no need
%for simulations. Moreover many results on hierarchical lattices closely follow
%what expected on Euclidean ones, thus showing that the physics
%on the formers is much the same as on the latters.
%We are thus confident that our analysis on hierarchical lattices sheds
%light on the non-universality problem on Euclidean ones.

\begin{figure}
\centerline{\psfig{file=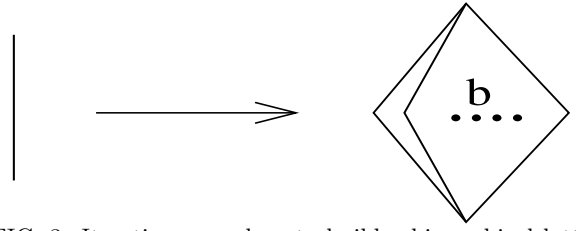,width=7.5cm,height=3cm,angle=0}}
\caption{Iterative procedure to build a hierarchical lattice: 
the bond on the left
schematically represents the lattice at step $n$; then it is used to build
the lattice at step $n+1$, represented on the right. In the picture only three
of the $b$ sides are explicitly drawn.}
\label{Fig: fig3}
\end{figure}

The main idea of hierarchical lattices is that it is possible to build
them iteratively, given a fundamental bond-block transformation, where a
bond is substituted by a block as in Fig.\ref{Fig: fig3}. The inverse process
can be seen as a coarse graining transformation. Indeed, using
this transformation, it can be shown that real-space renormalization
becomes exact on hierarchical lattices.
We exploit this property to write the corresponding renormalization for
the ground state of polymers on hierarchical lattices with
bond disorder taken from a given distribution.

%The renormalization procedure consists in following the evolution of 
%the probability distribution of the ground state energies.
Given the ground state energy distribution at a certain renormalization
step $n$, it is possible to compute the distribution
of the ground state energy at step $n+1$ via the
equation
\begin{equation}
P_{n+1}(x) = b Q_n(x) \left[ \int_x^\infty Q_n(x') dx' \right]^{b-1}
\label{basic equation}
\end{equation}
where $b$ is the number of sides as from Fig.\ref{Fig: fig3}.
$Q_n(x)$ is the convolution of two probability distributions $P_n(x)$,
$Q_n(x) = P_n(x) * P_n(x)$. The {\it r.h.s} of Eq.(\ref{basic equation})
represents the probability that one of the $b$ sides has an energy $x$,
while all the others have an energy greater than $x$. Due to the choice of the 
minimum energy, this is also the probability that the ground
state energy is $x$ at step $n+1$.

The number of sides $b$ can be related to a {\it fractal} dimension
of the lattice. Indeed, at every renormalization step we rescale
the length of the lattice of a factor $2$, and the volume of a factor 
$2b$. Therefore the dimension of the system can be
related to $b$ via the formula $D=1+\log_2 b$ (or, correspondingly,
$d= D-1 = \log_2 b$).

Given Eq.(\ref{basic equation}), the calculation of the
exponent $\omega$ reduces in principle to a numerical iteration 
of (\ref{basic equation}) given a starting distribution $P_0(x)$.
The ground state energy fluctuation exponent can be computed
as 
\begin{equation}
\omega_n = \log_2 \frac{\sigma_{n+1}}{\sigma_n}
\label{running omega}
\end{equation}
where $\sigma_n$ is the variance of $P_n(x)$.
Asymptotically, the running exponent $\omega_n$ 
tends to a constant $\omega_\infty$.
Possible non-universalities should then emerge using
different distributions. 
Yet, calculations of $\omega_\infty$ show that different distributions
give the same asymptotic values up to $b=20$
(corresponding to a substrate dimension $d=4.32...$).

As a byproduct, we see that hierarchical lattices 
of fractal dimension $D = d+1$ give values of the $\omega$
exponents close to the results on Euclidean lattices of the same dimension
($D=2$, $\omega_{hier} = 0.30(1)$, $\omega_{eucl} = 1/3$;
$D=3$, $\omega_{hier} = 0.22(1)$, $\omega_{eucl} = 0.24(1)$;
$D=4$, $\omega_{hier} = 0.15(1)$, $\omega_{eucl} = 0.16(1)$).

\begin{figure}
\centerline{\psfig{file=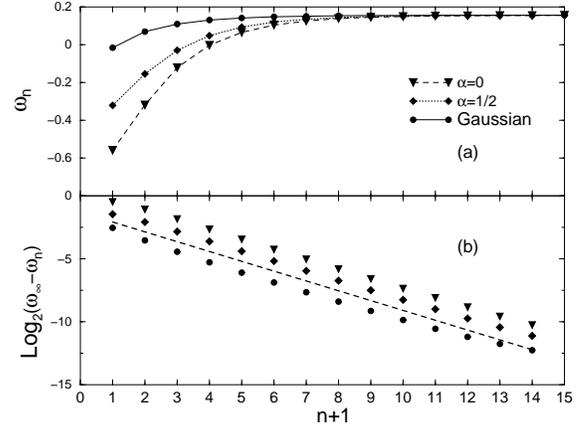,width=7.5cm,angle=0}}
\caption{a) Running exponents for $b=8$ with different disorder distributions,
namely Gaussian and as from Eq.(4); b) Log-log plot of the difference of the 
same running exponents from their asymptotic value, clearly showing the 
power-law behavior with $\gamma = 0.80(2)$. The value $b=8$ has been chosen
for numerical stability. }
\label{Fig: fig4}
\end{figure}

Although universality holds for directed polymers on hierarchical lattices,
we find that the running exponents $\omega_n$ show a power-law crossover toward
their asymptotic value $\omega_\infty$ (see Fig.\ref{Fig: fig4}a):
%This crossover depends indeed on the details of the microscopic 
%disorder distribution $P_0(x)$. Yet we find universality also in this 
%interesting crossover. As it is clear from Fig.\ref{Fig: fig4}b,
%there is a power-law approach of the running exponents to $\omega_\infty$
%
\begin{equation}
|\omega_\infty - \omega_n| \sim 2^{-\gamma (n+1)}\;\;,
\label{crossover}
\end{equation}
$2^{n+1}$ being the length of the lattice.

The power-law exponent $\gamma$ does not depend on the details of $P_0(x)$, 
that instead are responsible of the amplitude of the crossover.
We used Gaussian distributions and distributions of 
the form (\ref{swift distribution}),
as considered in \cite{NS97}.
%We see that the smaller is $\alpha$, the larger is the crossover
%amplitude as it appears from  Fig.\ref{Fig: fig4}a.
%This crossover is "slow": its power-law
%features (\ref{crossover}) 
%imply that there is no characteristic time after which
%the system can be truly considered in its asymptotic regime.

\begin{figure}
\centerline{\psfig{file=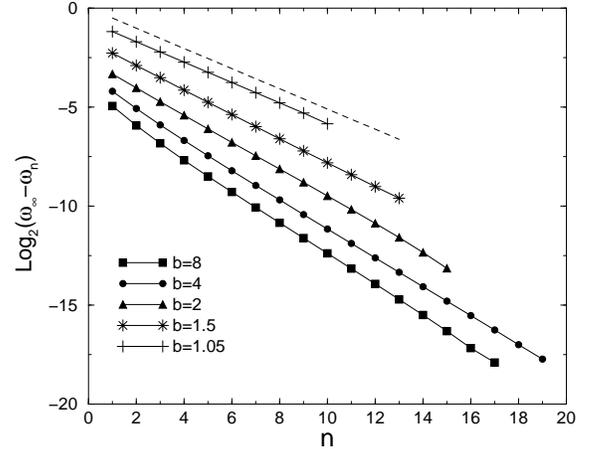,width=7.5cm,angle=0}}
\caption{Log-log plot of the difference of the running exponents from their
asymptotic value for different values of $b$. The dashed line represents
a power-law with exponent $-1/2$. All the data have been obtained using
Gaussian distributions.}
\label{Fig: fig5}
\end{figure}

In Fig.\ref{Fig: fig5} we show the running exponent approach to 
$\omega_\infty$ for different values of $b$ (all the numerics in this
case have been done using Gaussian distributions; as already explained above,
universality of the crossover exponent is verified).
We find that, as $b \to 1$ then $\gamma \to 1/2$.

We verify that indeed $\gamma(b=1^+) = 1/2$ 
via a $b=1+\epsilon$ expansion of Eq.(\ref{basic
equation}) on the same lines in Ref.\cite{DG89}: there a Gaussian 
microscopic distribution $P_0(x)$ was used, and no crossover effects were
discovered. We use instead
gamma distributions of the kind $P(x;\nu) = \frac{1}{\Gamma(\nu)}
x^{\nu-1} e^{-x}$, defined for $x > 0$: 
these distribution are {\it partially stable} under convolution;
indeed, $P(x;\nu) * P(x;\nu) = P(x; \nu' = 2\nu)$, that is, the convolution
of two gamma
distributions is still a gamma distribution 
(although characterized by a different exponent $\nu'$).

We assume that the distribution at iteration $n$ is given by 
\begin{equation}
P_n(x) = P_n^0(x) \left[ 1 + \epsilon \Phi_n(x) \right]
\label{perturbed distribution}
\end{equation}
where $P_n^0(x)$ is the distribution at iteration $n$ for the 
$1$-dimensional system ($b=1$), simply given by
$P_n^0(x) = P_{n-1}^0(x) * P_{n-1}^0(x) = Q_{n-1}^0(x)$.
The perturbation $\Phi_n(x)$ must satisfy the relation
\begin{equation}
\int_0^\infty \Phi_n(x) P_n^0(x) dx = 0\;\;.
\label{condition}
\end{equation}

We expand Eq.(\ref{basic equation}) for small $\epsilon$, keeping only terms 
of order $\epsilon$:
\begin{eqnarray}
Q_n^0(x) + &\epsilon& Q_n^0(x) \Phi_{n+1}(x) =
Q_n^0(x) + \nonumber \\
+ &\epsilon& \lbrace Q_n^0(x) + Q_n^0(x) \ln \int_x^\infty Q_n^0(x') dx' + 
 \nonumber \\
&+&  2 P_n^0(x) * \left[P_n^0(x) \Phi_n(x)\right] \rbrace.
\label{expansion}
\end{eqnarray}

The linear terms in 
$\epsilon$ are a recursion relation for the perturbation $\Phi(x)$,
\begin{eqnarray}
Q_n^0(x) \Phi_{n+1}(x) &=& Q_n^0(x) + Q_n^0(x) \ln \int_x^\infty Q_n^0(x') dx' 
+ \nonumber \\
&+& 2 P_n^0(x) * \left[P_n^0(x) \Phi_n(x)\right].
\label{recursion for phi}
\end{eqnarray}
All the terms in (\ref{recursion for phi}) are proportional to $Q_n^0(x)$
except for the last term on the {\it r.h.s}. We deal with this last term with
the {\it ansatz} equation
\begin{equation}
P_n^0(x) * \left[P_n^0(x) \phi_{n,s}(x)\right] = \lambda_{n,s} 
Q_n^0(x) \phi_{n+1,s}(x)
\label{eigenvalue equation}
\end{equation}
whose purpose is to extract from the {\it l.h.s.} expression a term 
proportional to  $Q_n^0(x)$ that can be then simplified 
in (\ref{recursion for phi}).
If among the solutions of Eq.(\ref{eigenvalue equation})
it is possible to find sets $\lbrace \phi_{n,s}\rbrace$ that are complete
and orthonormal
for any $n$, then we can write $\Phi_n(x) = \sum_s a_{n,s} \phi_{n,s}(x)$
and (\ref{recursion for phi}) becomes a recursion relation
for the $a_n$ coefficients.

%Before proceeding further we remark that, strictly speaking,
%Eq.(\ref{eigenvalue equation}) is not an eigenvalue equation.
%Indeed, it can be recast in an operatorial form as
%%
%\begin{equation}
%T_n \phi_{n,s} = \lambda_{n,s} \phi_{n+1,s}
%\label{operatorial form}
%\end{equation}
%%
%with $T_n$ a linear operator,
%showing that the $\phi$ functions on the {\it l.h.s.} and {\it r.h.s.}
%%of (\ref{operatorial form}) are relative to different iteration steps.
%This is not a major hindrance, since, as we will show, each set
%$\lbrace \phi_n \rbrace$ is orthonormal, and scalar products are
%taken only between functions belonging to the same iteration step.   
%
Any function of the form 
\begin{equation}
\phi_{n,s}(x) = \frac{(-1)^s}{\sqrt{s!}} 
\sqrt{\frac{\Gamma(\nu_n)}{\Gamma(\nu_n+s)}}
\frac{1}{P_n^0(x;\nu_n)} \frac{d^s}{dx^s} \left( x^s 
P_n^0(x;\nu_n) \right)
\label{eigenvectors}
\end{equation}
satisfies Eq.(\ref{eigenvalue equation}) with ``eigenvalues''
\begin{equation}
\lambda_{n,s} = \sqrt{\frac{\Gamma(\nu_n+s) \Gamma(\nu_{n+1})}
{\Gamma(\nu_n) \Gamma(2 \nu_n+s)}}.
\label{eigenvalues}
\end{equation}
Eq.(\ref{eigenvectors}) is nothing else than Rodrigues' formula for the 
generalized Laguerre polynomials, properly normalized, that
are known to be orthogonal and complete with weight $P_n^0(x)$\cite{gaussian}.
The first three such polynomials are 
\begin{eqnarray}
\phi_{n,0}(x) &=& 1\;, \nonumber \\
\phi_{n,1}(x) &=& (x-\nu_n)/\sqrt{\nu_n}\;, \nonumber \\
\phi_{n,2}(x) &=& (x^2 - 2(\nu_n+1) x + \nu_n (\nu_n+1))/
\sqrt{2 \nu_n (\nu_n+1)}\;. \nonumber \\
&& 
\label{first three}
\end{eqnarray}
Substituting the series expansion of $\Phi_n(x)$ 
in (\ref{recursion for phi}),
it is possible to read the iteration equations for the 
coefficients $a_{n,s}$:
\begin{eqnarray}
a_{n+1,0} & = & 1 + K_{n,0} + 2 \lambda_{n,0} a_{n,0} \nonumber \\
a_{n+1,s} & = & K_{n,s} + 2 \lambda_{n,s} a_{n,s} \;\;\;\; s > 0
\label{equations for a}
\end{eqnarray}
with
\begin{equation}
K_{n,s} = \int_0^\infty dx P_{n+1}^0(x) \phi_{n+1,s} \ln \int_x^\infty 
P_{n+1}^0(x') dx'
\label{K}
\end{equation}
It is straightforward to show that $K_0 = -1$; moreover from Eq.
(\ref{condition}) we find
$a_{0,0} = 0$ and therefore $a_{n,0} = 0$ for any $n$, as required.
It is also possible to show that $2 \lambda_{n,s} < 1$ if $s>2$
(indeed the leading behavior for large $\nu_n$, that is for large 
$n$, is $2 \lambda_{n,s} \sim 2^{1-s/2} [1+(s-1/2) s/2\nu_n]^{1/2}$).
Therefore all the iterations equations (\ref{equations for a}) with 
$s>2$ converge and are asymptotically irrelevant for
the scaling behavior. We are left with $s=1,2$.
Indeed $a_{n,1}$ and $a_{n,2}$ are the only coefficients that
are relevant for the computation of $\omega_n$ from (\ref{running omega}).
%
%The average energy can be written as
%%
%\begin{equation}
%<x>_n = \int_0^\infty x P_n(x) dx = \nu_n + \epsilon \sqrt{\nu_n} a_{n,1}
%\label{average}
%\end{equation}
%%
%and the second moment is
%%
%\begin{eqnarray}
%<x^2>_n &=& \int_0^\infty x^2 P_n(x) dx = \nu_n (\nu_n+1) + \nonumber \\
%&+&  \epsilon (\sqrt{2 \nu_n (\nu_n+1)} a_{n,2} + 2 (\nu_n + 1) \sqrt{\nu_n} a_{n,1})
%\label{second moment}
%\end{eqnarray}
%%

We can then compute the variance $\sigma_n^2$ as
\begin{eqnarray}
\sigma_n^2 &=& <x^2>_n - <x>_n^2 = \nonumber \\
&=& \nu_n + \epsilon 
(\sqrt{2 \nu_n (\nu_n+1)} a_{n,2} + 2 \sqrt{\nu_n} a_{n,1})
\label{variance}
\end{eqnarray}

We can write then 
\begin{equation}
\frac{\sigma_{n+1}^2}{\sigma_n^2} = 2 \left[1+\epsilon \left(\sqrt{2} K_{n,2}
+ 2 \frac{K_{n,1}}{\sqrt{\nu_{n+1}}}\right) \right]
\label{variance ratio}
\end{equation}
where we keep only the leading and next-to-the-leading terms.
From (\ref{variance ratio}) the 
order $\epsilon$ correction to the $\omega$ exponent can be computed.
Indeed, $\sigma_{n+1}^2/\sigma_n^2 = 2^{2\omega_n}$,
and since $\nu_{n+1} = 2^{n+1} \nu_0$, we can write 
\begin{equation}
\omega_n = \frac{1}{2} + \epsilon \frac{\sqrt{2}}{\ln 2} K_{n,2} + \epsilon
\frac{2}{\ln 2 \sqrt{\nu_0}} K_{n,1} 2^{-\frac{n+1}{2}}.
\label{power 1/2}
\end{equation}
Furthermore, both $K_{n,1}$ and $K_{n,2}$ have a power-law convergence
to their asymptotic value, with exponent $1/2$
%\cite{approx} 
(this
result can be obtained using a Gaussian approximation in their evaluation). 
As a result
the {\it r.h.s.} of  (\ref{power 1/2}) converges
with exponent $\gamma = 1/2$ to $\omega_\infty$.

In conclusion, we have found numerically that both the KPZ
equation on $d=1$ Euclidean lattices and DPRM on hierarchical lattices
show universal behavior, being the exponent
$\omega$ independent on the details of the disorder distribution.
Yet, the presence of non trivial 
power-law (therefore "slow") crossover effects has been unveiled.
%These crossovers are governed by a power-law approach to 
%the asymptotic regime: they are therefore "slow" crossovers,
%since the approach does not show any characteristic time.
Moreover, we have shown that the crossover exponent $\gamma$ is 
universal. 
%not depending on the details of the
%chosen microscopic disorder distribution, but only on the dimensionality
%of the system. 
%It depends only on the
%dimensionality of the system (related to the number $b$ of sides of
%the lattice), 
We have also shown numerically and analytically
that $\gamma \to 1/2$ when $b \to 1$.

Since we find the same qualitative slow crossover behavior both on $d=1$
Euclidean lattices and on hierarchical lattices of any dimension, 
it is reasonable to think that the same behavior is present also for
$d \ge 2$ Euclidean lattices, providing a way out
of the non-universality claimed in~\cite{NS97}.

We thank C. Tebaldi for useful discussions.

\begin{thebibliography}{99}

\bibitem{HH85+}
D.A. Huse and C. Henley, Phys. Rev. Lett. {\bf 54}, 2708 (1985);
D.A. Huse, C. Henley and D.S. Fisher, Phys. Rev. Lett. {\bf 55}, 2924 (1985). 

\bibitem{KPZ86+}
M. Kardar, G. Parisi and Y.-C. Zhang, Phys. Rev. Lett. {\bf 56}, 889 (1986);
M. Marsili, A. Maritan, F. Toigo and J.R. Banavar, Rev. Mod. Phys. {\bf 68},
963 (1996).

\bibitem{HHZ95} T. Halpin-Healy and Y.-C. Zhang, Phys. Rep. {\bf 254},
215 (1995), and references therein.

\bibitem{NS97} T.J. Newman and M. Swift, Phys. Rev. Lett. {\bf 79},
2261 (1997).

\bibitem{DG89} B. Derrida and R.B. Griffiths, Europhys. Lett. {\bf 8},
111 (1989).

\bibitem{gaussian} All the discussion from Eq.(\ref{eigenvalue equation}) to 
Eq.(\ref{eigenvalues}) can be repeated step-by-step
using Gaussian distributions. In that case the good $\lbrace \phi_{n} \rbrace$
functions are Hermite polynomials. This result was obtained also
in \cite{DG89}, although in an implicit and not generalizable form.

%\bibitem{approx} This result can be shown using a Gaussian 
%approximation on the $K_{n,1}$ and $K_{n,2}$ integrals. Indeed, $P_n^0(x)$
%tends to a Gaussian distribution as $n$ increases due to the central 
%limit theorem. Expanding then $\phi_{n,1}(x)$
%and $\phi_{n,2}(x)$ in Hermite polynomials 
%it is possible to obtain the asymptotic value of the integrals 
%and the power-law decaying corrections\cite{DG89,gaussian}.
%

\end{thebibliography}
\end{document}